\documentclass[final]{aipproc}
\usepackage{amssymb}

\layoutstyle{8x11single}

\begin{document}

\title{The canonical Gamma-Ray Bursts and their ``precursors''}


\author{Remo Ruffini}{address={ICRANet, Piazzale della Repubblica 10, 65122 Pescara, Italy.}, altaddress={Dipartimento di Fisica, Universit\`a di Roma ``La Sapienza'', P.le Aldo Moro 5, 00185 Roma, Italy.}}

\author{Alexey G. Aksenov}{address={Institute for Theoretical and Experimental Physics, B. Cheremushkinskaya, 25, 117218 Moscow, Russia}
}

\author{Maria Grazia Bernardini}{address={ICRANet, Piazzale della Repubblica 10, 65122 Pescara, Italy.}, altaddress={Dipartimento di Fisica, Universit\`a di Roma ``La Sapienza'', P.le Aldo Moro 5, 00185 Roma, Italy.}}

\author{Carlo Luciano Bianco}{address={ICRANet, Piazzale della Repubblica 10, 65122 Pescara, Italy.}, altaddress={Dipartimento di Fisica, Universit\`a di Roma ``La Sapienza'', P.le Aldo Moro 5, 00185 Roma, Italy.}}

\author{Letizia Caito}{address={ICRANet, Piazzale della Repubblica 10, 65122 Pescara, Italy.}, altaddress={Dipartimento di Fisica, Universit\`a di Roma ``La Sapienza'', P.le Aldo Moro 5, 00185 Roma, Italy.}}

\author{Maria Giovanna Dainotti}{address={ICRANet, Piazzale della Repubblica 10, 65122 Pescara, Italy.}, altaddress={Dipartimento di Fisica, Universit\`a di Roma ``La Sapienza'', P.le Aldo Moro 5, 00185 Roma, Italy.}}

\author{Gustavo De Barros}{address={ICRANet, Piazzale della Repubblica 10, 65122 Pescara, Italy.}, altaddress={Dipartimento di Fisica, Universit\`a di Roma ``La Sapienza'', P.le Aldo Moro 5, 00185 Roma, Italy.}}

\author{Roberto Guida}{address={ICRANet, Piazzale della Repubblica 10, 65122 Pescara, Italy.}, altaddress={Dipartimento di Fisica, Universit\`a di Roma ``La Sapienza'', P.le Aldo Moro 5, 00185 Roma, Italy.}}

\author{Gregory V. Vereshchagin}{address={ICRANet, Piazzale della Repubblica 10, 65122 Pescara, Italy.}}

\author{She-Sheng Xue}{address={ICRANet, Piazzale della Repubblica 10, 65122 Pescara, Italy.}}

\begin{abstract}
The fireshell model for Gamma-Ray Bursts (GRBs) naturally leads to a canonical GRB composed of a proper-GRB (P-GRB) and an afterglow. P-GRBs, introduced by us in 2001, are sometimes considered ``precursors'' of the main GRB event in the current literature. We show in this paper how the fireshell model leads to the understanding of the structure of GRBs, with precise estimates of the time sequence and intensities of the P-GRB and the of the afterglow. It leads as well to a natural classification of the canonical GRBs which overcomes the traditional one in short and long GRBs.
\end{abstract}

\maketitle

The so-called ``prompt emission'' light curves of many Gamma-Ray Bursts (GRBs) present a small pulse preceding the main GRB event, with a lower peak flux and separated by this last one by a quiescent time. The nature of such GRB ``precursors'' is one of the most debated issues in the current literature \citep[see Ref.][as well as G. Ghisellini's talk at this Meeting]{2008arXiv0806.3076B}. Already in 2001 \citep{2001ApJ...555L.113R}, within the ``fireshell'' model, we proposed that GRB ``precursors'' are the Proper GRBs (P-GRBs) emitted when the fireshell becomes transparent, and we gave precise estimates of the time sequence and intensities of the P-GRB and the of the afterglow.

Within our approach, in fact, we assume that all GRBs originate from the gravitational collapse to a black hole \citep{2001ApJ...555L.113R,2007AIPC..910...55R}. The $e^\pm$ plasma created in the process of the black hole formation reaches thermal equilibrium on a time scale on the order of $\sim 10^{-13}$ s \citep{2007PhRvL..99l5003A}. Then it expands as an optically thick and spherically symmetric ``fireshell'' with a constant width in the laboratory frame, i.e. the frame in which the black hole is at rest \citep{1999A&A...350..334R}. During its expansion, the fireshell engulfs the surrounding baryonic remnants \citep{2000A&A...359..855R}. This optically thick self-acceleration phase of the fireshell lasts until the transparency condition is reached and the Proper-GRB (P-GRB) is emitted \citep{2001ApJ...555L.113R,2007AIPC..910...55R}. This phase is fully determined by the only two free parameters characterizing the source \citep{2001ApJ...555L.113R,2007AIPC..910...55R}, namely the total initial energy $E_{e^\pm}^{tot}$ of the $e^\pm$ plasma and the fireshell engulfed baryon loading $B\equiv M_Bc^2/E_{e^\pm}^{tot}$, where $M_B$ is the total baryons' mass \citep{2000A&A...359..855R}. For the fireshell dynamics to reach ultrarelativistic velocities after the baryonic matter engulfment it must be $B < 10^{-2}$ \citep{2000A&A...359..855R}. After the transparency condition is reached and the P-GRB has been emitted, it remains an accelerated optically thin fireshell formed only of baryons, which is ballistically expanding into the CircumBurst Medium (CBM). Such a baryonic shell progressively loses its kinetic energy interacting with the CBM, and this gives rise to the ``afterglow'' emission, which comprises a rising part, a peak and a decaying tail \citep[see also Bianco et al. contribution in this volume]{2001ApJ...555L.113R,2007AIPC..910...55R}.

We define a ``canonical GRB'' light curve with two sharply different components: the P-GRB and the afterglow. Their relative energetics and time separation are functions of $E_{e^\pm}^{tot}$ and $B$ \citep[][see also Bianco et al. contribution in this volume]{2001ApJ...555L.113R,2006ApJ...645L.109R,2007AIPC..910...55R,2008AIPC.1000..305B}. In this respect, in the current literature \citep[see e.g. Refs.][and references therein]{2005RvMP...76.1143P,2006RPPh...69.2259M} there is somewhat a state of confusion: what is usually called the ``prompt emission'' comprises the P-GRB, the rising part and the peak of the afterglow. Similarly, what is usually called the ``afterglow'' is just its decaying tail.

\begin{figure}
\centering
\includegraphics[width=0.76\hsize]{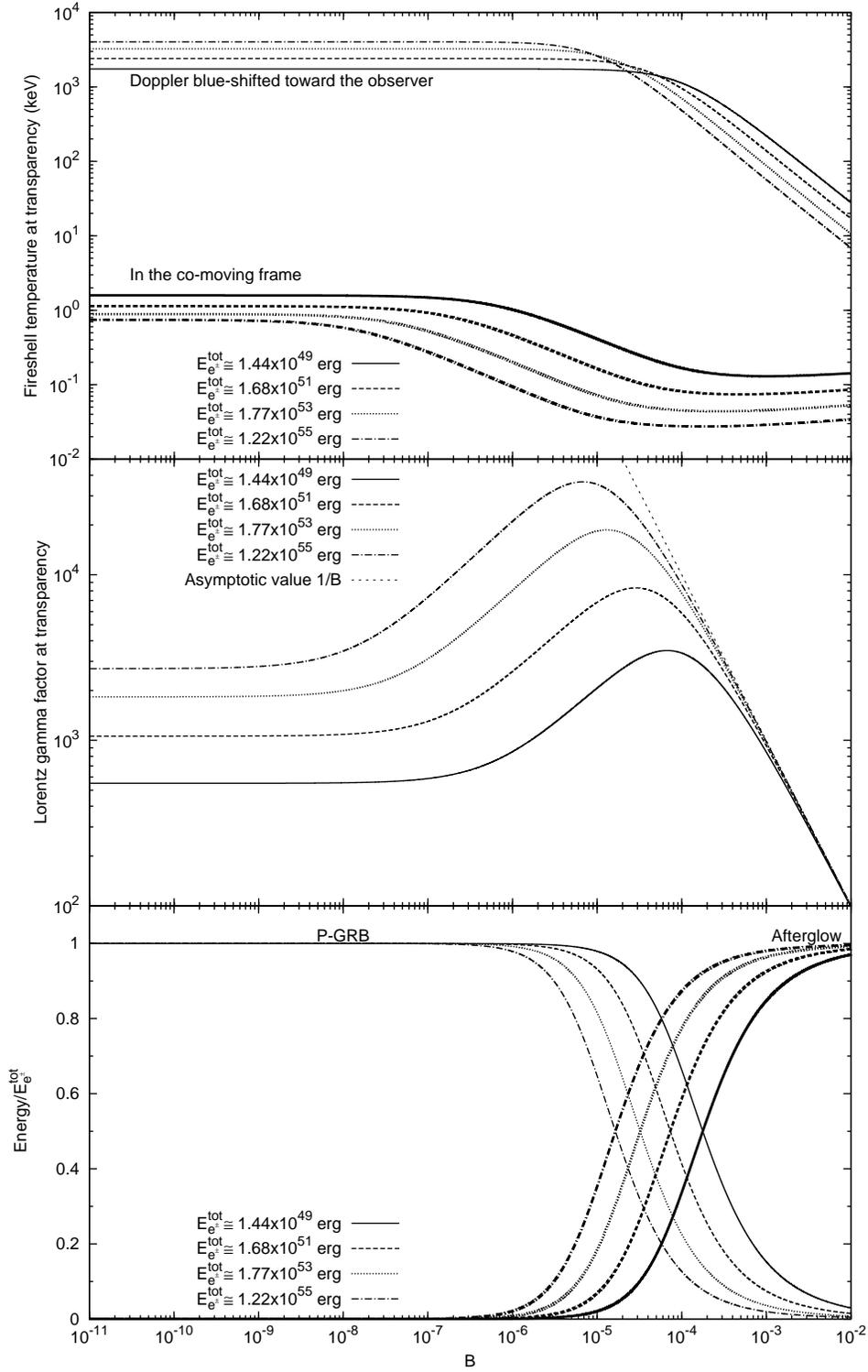}
\caption{At the fireshell transparency point, for $4$ different values of $E^{tot}_{e^\pm}$, we plot as a function of $B$: \textbf{(Above)} The fireshell temperature in the co-moving frame $T_\circ^{com}$ (ticker lines) and the one Doppler blue-shifted along the line of sight toward the observer in the source cosmological rest frame $T_\circ^{obs}$ (thinner lines); \textbf{(Middle)} The fireshell Lorentz gamma factor $\gamma_\circ$ together with the asymptotic value $\gamma_\circ = 1/B$; \textbf{(Below)} The energy radiated in the P-GRB (thinner lines, rising when $B$ decreases) and the one converted into baryonic kinetic energy and later emitted in the afterglow (thicker lines, rising when $B$ increases), in units of $E_{e^\pm}^{tot}$.}
\label{ftemp-fgamma-bcross}
\end{figure}

We have recently shown \citep{PRDsub} that a thermal spectrum still occurs in presence of $e^\pm$ pairs and baryons. By solving the rate equation we have evaluated the evolution of the temperature during the fireshell expansion, all the way up to when the transparency condition is reached \citep{1999A&A...350..334R,2000A&A...359..855R}. In the upper panel of Fig.~\ref{ftemp-fgamma-bcross} we plot, as a function of $B$, the fireshell temperature $T_\circ$ at the transparency point, i.e. the temperature of the P-GRB radiation. The plot is drawn for four different values of $E_{e^\pm}^{tot}$ in the interval $[10^{49}, 10^{55}]$ ergs, well encompassing GRBs' observed isotropic energies. We plot both the value in the co-moving frame $T_\circ^{com}$ and the one Doppler blue-shifted toward the observed $T_\circ^{obs} = (1+\beta_\circ) \gamma_\circ T_\circ^{com}$, where $\beta_\circ$ is the fireshell speed at the transparency point in units of $c$ \citep{2000A&A...359..855R}.

In the middle panel of Fig.~\ref{ftemp-fgamma-bcross} we plot, as a function of $B$, the fireshell Lorentz gamma factor at the transparency point $\gamma_\circ$. The plot is drawn for the same four different values of $E_{e^\pm}^{tot}$ of the upper panel. Also plotted is the asymptotic value $\gamma_\circ = 1/B$, which corresponds to the condition when the entire initial internal energy of the plasma $E_{e^\pm}^{tot}$ has been converted into kinetic energy of the baryons \citep{2000A&A...359..855R}. We see that such an asymptotic value is approached for $B \to 10^{-2}$. We see also that, if $E_{e^\pm}^{tot}$ increases, the maximum values of $\gamma_\circ$ are higher and they are reached for lower values of $B$.

In the lower panel of Fig.~\ref{ftemp-fgamma-bcross} we plot, as a function of $B$, the total energy radiated at the transparency point in the P-GRB and the one converted into baryonic kinetic energy and later emitted in the afterglow. The plot is drawn for the same four different values of $E_{e^\pm}^{tot}$ of the upper panel. We see that for $B \lesssim 10^{-5}$ the total energy emitted in the P-GRB is always larger than the one emitted in the afterglow: we have what we call a ``genuine'' short GRB \citep[][see also Bianco et al. contribution in this volume]{2001ApJ...555L.113R,2007A&A...474L..13B,2008AIPC.1000..305B}. On the other hand, for $3.0\times 10^{-4} \lesssim B < 10^{-2}$ the total energy emitted in the P-GRB is always smaller than the one emitted in the afterglow. If it is not below the instrumental threshold, the P-GRB can be observed in this case as a small pulse preceding the main GRB event (which coincides with the peak of the afterglow), i.e. as a GRB ``precursor'' \citep{2001ApJ...555L.113R,2003AIPC..668...16R,2008AIPC.1000..305B}. The only exception to this rule is when we are in presence of a peculiarly small CBM density which ``deflates'' the afterglow peak luminosity with respect to the P-GRB one and gives rise to a ``disguised'' short GRB \citep[][see also Bianco et al. contribution in this volume]{2007A&A...474L..13B,2008AIPC.1000..305B}. Finally, for $10^{-5} \lesssim B \lesssim 3.0\times 10^{-4}$ we are in an intermediate case, and the energetic predominance of the P-GRB over the afterglow depends on the value of $E_{e^\pm}^{tot}$.

Particularly relevant for the new era of the \emph{Agile} and \emph{GLAST} satellites is that for $B < 10^{-3}$ the P-GRB emission has an observed temperature up to $10^{3}$ keV or higher. This high-energy emission has been unobservable by the \emph{Swift} satellite.

\begin{figure}
\centering
\includegraphics[width=0.71\hsize]{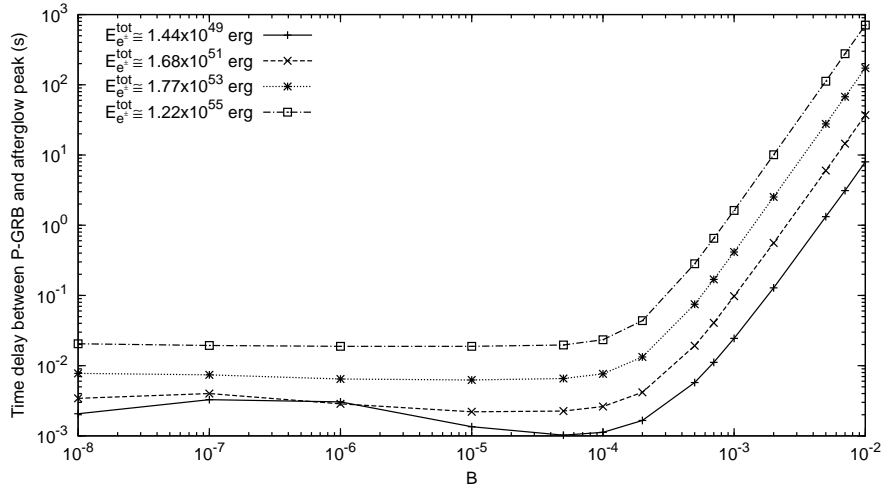}
\caption{For $4$ different values of $E^{tot}_{e^\pm}$, we plot as a function of $B$ the arrival time separation $\Delta t_a$ between the P-GRB and the peak of the afterglow (i.e. the ``quiescent time between the ``precursor'' and the main GRB event), measured in the source cosmological rest frame. This computation has been performed assuming a constant CBM density $n_{cbm}=1.0$ particles/cm$^3$. The points represents the actually numerically computed values, connected by straight line segments.}
\label{dta}
\end{figure}

In Fig.~\ref{dta} we plot, as a function of $B$, the arrival time separation $\Delta t_a$ between the P-GRB and the peak of the afterglow measured in the cosmological rest frame of the source. Such a time separation $\Delta t_a$ is the ``quiescent time'' between the precursor (i.e. the P-GRB) and the main GRB event (i.e. the peak of the afterglow). The plot is drawn for the same four different values of $E_{e^\pm}^{tot}$ of Fig.~\ref{ftemp-fgamma-bcross}. The arrival time of the peak of the afterglow emission depends on the detailed profile of the CBM density. In this plot it has been assumed a constant CBM density $n_{cbm}=1.0$ particles/cm$^3$. We can see that, for $3.0\times 10^{-4} \lesssim B < 10^{-2}$, which is the condition for P-GRBs to be ``precursors'' (see above), $\Delta t_a$ increases both with $B$ and with $E_{e^\pm}^{tot}$. We can have $\Delta t_a > 10^2$ s and, in some extreme cases even $\Delta t_a \sim 10^3$ s. For $B \lesssim 3.0\times 10^{-4}$, instead, $\Delta t_a$ presents a behavior which qualitatively follows the opposite of $\gamma_\circ$ (see middle panel of Fig.~\ref{ftemp-fgamma-bcross}).

\begin{figure}
\centering
\includegraphics[width=0.71\hsize]{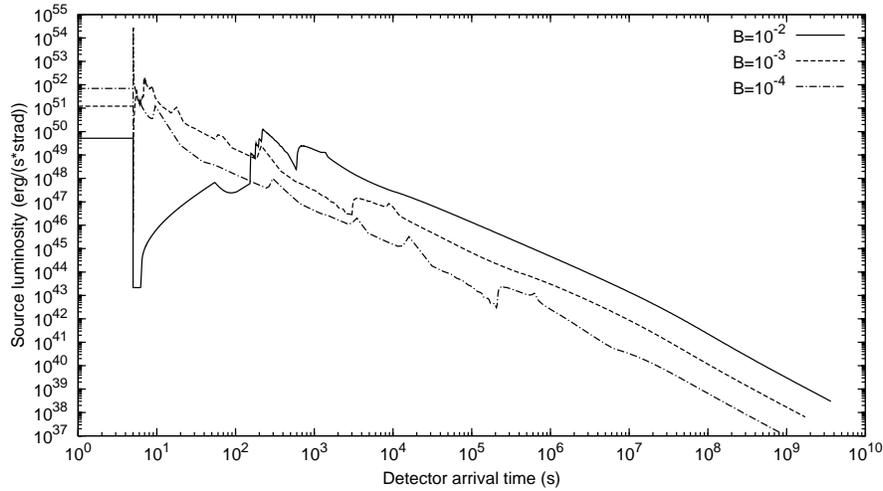}
\caption{We plot three theoretical afterglow bolometric light curves together with the corresponding P-GRB peak luminosities (the horizontal segments). The computations have been performed assuming the same $E^{tot}_{e^\pm}$ and CBM structure of GRB991216 and three different values of $B$. The P-GRBs have been assumed to have the same duration in the three cases, i.e. $5$ s. For $B$ decreasing, the afterglow light curve squeezes itself on the P-GRB.}
\label{multi_b}
\end{figure}

Finally, in Fig.~\ref{multi_b} we present three theoretical afterglow bolometric light curves together with the corresponding P-GRB peak luminosities for three different values of $B$. The duration of the P-GRBs has been assumed to be the same in the three cases (i.e. $5$ s). The computations have been performed assuming the same $E^{tot}_{e^\pm}$ and the same detailed CBM density profile of GRB991216 \citep{2003AIPC..668...16R}. In this picture we clearly see how, for $B$ decreasing, the afterglow light curve ``squeezes'' itself on the P-GRB and the P-GRB peak luminosity increases.

Before closing, we like to mention that, using the diagrams represented in Figs.~\ref{ftemp-fgamma-bcross}-\ref{dta}, in principle one can compute the two free parameters of the fireshell model, namely $E^{tot}_{e^\pm}$ and $B$, from the ratio between the total energies of the P-GRB and of the afterglow and from the temporal separation between the peaks of the corresponding bolometric light curves. None of these quantities depends on the cosmological model. Therefore, one can in principle use this method to compute the GRBs' intrinsic luminosity and make GRBs the best cosmological distance indicators available today. The increase of the number of observed sources, as well as the more accurate knowledge of their CBM density profiles, will possibly make viable this procedure to test cosmological parameters, in addition to the Amati relation \citep{2008arXiv0805.0377A,2008A&A...487L..37G}.


\begin{thebibliography}{15}
\expandafter\ifx\csname natexlab\endcsname\relax\def\natexlab#1{#1}\fi
\providecommand{\enquote}[1]{``#1''}
\expandafter\ifx\csname url\endcsname\relax
  \def\url#1{\texttt{#1}}\fi
\expandafter\ifx\csname urlprefix\endcsname\relax\def\urlprefix{URL }\fi
\providecommand{\eprint}[2][]{\url{#2}}

\bibitem[{Burlon} et~al.(in press)]{2008arXiv0806.3076B}
D.~{Burlon}, G.~{Ghirlanda}, G.~{Ghisellini}, D.~{Lazzati}, L.~{Nava},
  M.~{Nardini}, and A.~{Celotti}, \emph{ApJ}  (in press),
  \eprint{arXiv:0806.3076}.

\bibitem[{Ruffini} et~al.(2001)]{2001ApJ...555L.113R}
R.~{Ruffini}, C.~L. {Bianco}, P.~{Chardonnet}, F.~{Fraschetti}, and S.-S.
  {Xue}, \emph{ApJ} \textbf{555}, L113--L116 (2001).

\bibitem[{Ruffini} et~al.(2007)]{2007AIPC..910...55R}
R.~{Ruffini}, M.~G. {Bernardini}, C.~L. {Bianco}, L.~{Caito}, P.~{Chardonnet},
  M.~G. {Dainotti}, F.~{Fraschetti}, R.~{Guida}, M.~{Rotondo},
  G.~{Vereshchagin}, L.~{Vitagliano}, and S.-S. {Xue}, \enquote{The Blackholic
  energy and the canonical Gamma-Ray Burst,} in \emph{XIIth Brazilian School of
  Cosmology and Gravitation}, edited by M.~{Novello}, and S.~E. {Perez
  Bergliaffa}, 2007, vol. 910 of \emph{American Institute of Physics Conference
  Series}, pp. 55--217.

\bibitem[{Aksenov} et~al.(2007)]{2007PhRvL..99l5003A}
A.~{Aksenov}, R.~{Ruffini}, and G.~{Vereshchagin}, \emph{Phys. Rev. Lett.}
  \textbf{99}, 125003 (2007).

\bibitem[{Ruffini} et~al.(1999)]{1999A&A...350..334R}
R.~{Ruffini}, J.~D. {Salmonson}, J.~R. {Wilson}, and S.-S. {Xue}, \emph{A\&A}
  \textbf{350}, 334--343 (1999).

\bibitem[{Ruffini} et~al.(2000)]{2000A&A...359..855R}
R.~{Ruffini}, J.~D. {Salmonson}, J.~R. {Wilson}, and S.-S. {Xue}, \emph{A\&A}
  \textbf{359}, 855--864 (2000).

\bibitem[{Ruffini} et~al.(2006)]{2006ApJ...645L.109R}
R.~{Ruffini}, M.~G. {Bernardini}, C.~L. {Bianco}, P.~{Chardonnet},
  F.~{Fraschetti}, R.~{Guida}, and S.-S. {Xue}, \emph{ApJ} \textbf{645},
  L109--L112 (2006).

\bibitem[{Bianco} et~al.(2008)]{2008AIPC.1000..305B}
C.~L. {Bianco}, M.~G. {Bernardini}, L.~{Caito}, M.~G. {Dainotti}, R.~{Guida},
  and R.~{Ruffini}, \enquote{Short and canonical GRBs,} in \emph{GAMMA-RAY
  BURSTS 2007: Proceedings of the Santa Fe Conference}, edited by M.~{Galassi},
  D.~{Palmer}, and E.~{Fenimore}, 2008, vol. 1000 of \emph{American Institute
  of Physics Conference Series}, pp. 305--308.

\bibitem[{Piran}(2005)]{2005RvMP...76.1143P}
T.~{Piran}, \emph{Reviews of Modern Physics} \textbf{76}, 1143--1210 (2005).

\bibitem[{Meszaros}(2006)]{2006RPPh...69.2259M}
P.~{Meszaros}, \emph{Reports of Progress in Physics} \textbf{69}, 2259--2322
  (2006).

\bibitem[{Aksenov} et~al.(submitted to)]{PRDsub}
A.~{Aksenov}, R.~{Ruffini}, and G.~{Vereshchagin}, \emph{Phys. Rev. D}
  (submitted to).

\bibitem[{Bernardini} et~al.(2007)]{2007A&A...474L..13B}
M.~G. {Bernardini}, C.~L. {Bianco}, L.~{Caito}, M.~G. {Dainotti}, R.~{Guida},
  and R.~{Ruffini}, \emph{A\&A} \textbf{474}, L13--L16 (2007).

\bibitem[{Ruffini} et~al.(2003)]{2003AIPC..668...16R}
R.~{Ruffini}, C.~L. {Bianco}, P.~{Chardonnet}, F.~{Fraschetti},
  L.~{Vitagliano}, and S.-S. {Xue}, \enquote{New perspectives in physics and
  astrophysics from the theoretical understanding of Gamma-Ray Bursts,} in
  \emph{Cosmology and Gravitation}, edited by M.~{Novello}, and S.~E. {Perez
  Bergliaffa}, 2003, vol. 668 of \emph{American Institute of Physics Conference
  Series}, pp. 16--107.

\bibitem[{Amati} et~al.(submitted to)]{2008arXiv0805.0377A}
L.~{Amati}, C.~{Guidorzi}, F.~{Frontera}, M.~{Della Valle}, F.~{Finelli},
  R.~{Landi}, and E.~{Montanari}, \emph{MNRAS}  (submitted to),
  \eprint{arXiv:0805.0377}.

\bibitem[{Guida} et~al.(2008)]{2008A&A...487L..37G}
R.~{Guida}, M.~G. {Bernardini}, C.~L. {Bianco}, L.~{Caito}, M.~G. {Dainotti},
  and R.~{Ruffini}, \emph{A\&A} \textbf{487}, L37--L40 (2008).

\end{thebibliography}
\end{document}